# Intrinsic Phase Diagram of Superconductivity in the BiCh₂-based System Without In-plane Disorder

Kouhei Nagasaka[1], Atsuhiro Nishida[1], Rajveer Jha[2], Joe Kajitani[2], Osuke Miura[1], Ryuji Higashinaka[2], Tatsuma D. Matsuda[2], Yuji Aoki[2], Akira Miura[3], Chikako Moriyoshi[4], Yoshihiro Kuroiwa[4], Hidetomo Usui[5], Kazuhiko Kuroki[5], Yoshikazu Mizuguchi[1*]

[1]*Department of Electrical and Electronic Engineering, Tokyo Metropolitan University, Hachioji 192-0397, Japan*

[2]*Department of Physics, Tokyo Metropolitan University, Hachioji 192-0397, Japan*

[3]*Faculty of Engineering, Hokkaido University, Kita-13, Nishi-8, Kita-ku, Sapporo, Hokkaido 060-8628, Japan*

[4]*Department of Physical Science, Hiroshima University, 1-3-1 Kagamiyama, Higashihiroshima, Hiroshima 739-8526, Japan*

[5]*Department of Physics, Osaka University, Machikaneyama-cho, Toyonaka 560-0043, Japan*

We have investigated the crystal structure and physical properties of LaO$_{1-x}$F$_x$BiSSe to reveal the intrinsic superconductivity phase diagram of the BiCh₂-based layered compound family. From synchrotron X-ray diffraction and Rietveld refinements with anisotropic displacement parameters, we clearly found that the in-plane disorder in the BiSSe layer was fully suppressed for all $x$. In LaO$_{1-x}$F$_x$BiSSe, metallic conductivity and superconductivity are suddenly induced by electron doping even at $x = 0.05$ and 0.1 with a monoclinic structure. In addition, $x$ (F concentration) dependence of the transition temperature ($T_c$) for $x = 0.2$-0.5 with a tetragonal structure shows an anomalously flat phase diagram. With these experimental facts, we have proposed the intrinsic phase diagram of the ideal BiCh₂-based superconductors with less in-plane disorder.

Since the discovery of Bi$_4$O$_4$S$_3$ and REO$_{1-x}$F$_x$BiS$_2$ (RE: rare earth) superconductors, BiCh₂-based (Ch: chalcogen) superconductors have been drawing much attention as a new class of layered superconductors [1-3]. Since the crystal structure composed of alternate stacks of the electrically conducting BiCh₂ layers and the insulating (blocking) layers resembles those of the Cuprate and the FeAs-based high-transition-temperature (high-$T_c$) superconductors [4,5], many experimental and theoretical studies have been performed to clarify the superconductivity mechanisms of this system and to increase $T_c$. However, the mechanisms have not been understood completely. Recently, Morice *et al.* proposed that a weak-coupling electron-phonon mechanism cannot explain the emerging $T_c$, as high as 11 K, in the $BiCh_2$-based systems, from first principle calculations [6]. Therefore, full understandings of the basic characteristics of the superconductivity in the $BiCh_2$-based system are crucial.

The parent phase of the $BiCh_2$-based superconductor is a band insulator [1,7,8]. On the basis of the calculated band structure, electrons carriers are doped into the bands mainly composed of Bi-$6p_x$ and Bi-$6p_y$ components. Electron-doped $BiCh_2$-based compounds are expected to become metallic. Indeed, superconductivity is experimentally observed in electron-doped compounds [1-3]. However, the real situations are not simple as expected from the band structure. Although the superconductivity in $BiCh_2$-based compounds is emerged by carrier doping, metallic transport is sometimes absent, and weakly localized behavior (semiconducting-like behavior) is observed in electrical resistivity measurements; a good example would be optimally doped $LaO_{0.5}F_{0.5}BiS_2$. In semiconducting-like samples of $LaO_{0.5}F_{0.5}BiS_2$, bulk superconductivity is not observed, while weak (filamentary) superconductivity is observed. To induce bulk superconductivity in the $LaO_{0.5}F_{0.5}BiS_2$ system, external pressure effects [9-16] and/or element substitution at the La site [17-20], which optimize the crystal structure, are available. Namely, both electron carrier doping and crystal structure optimization are required for the emergence of bulk superconductivity in the $BiCh_2$-based system. Recently, we proposed that the enhancement of in-plane chemical pressure is the factor essential for the $BiCh_2$-based superconductivity [21]. The in-plane chemical pressure is related to the orbital overlaps between Bi and Ch and can be tuned by element substitutions at the blocking layer or the conducting layer. For example, in $REO_{0.5}F_{0.5}BiS_2$, isovalent substitution of $La^{3+}$ (at the RE site) by smaller $Pr^{3+}$, $Nd^{3+}$, or $Sm^{3+}$ induces bulk superconductivity [17-20], and $T_c$ reaches above 5 K in $NdO_{0.5}F_{0.5}BiS_2$ or $Nd_{1-x}Sm_xO_{0.5}F_{0.5}BiS_2$ [17,22]. Basically, the $RE^{3+}$ substitution does not affect the structural symmetry but compresses Bi-S planes along the in-plane direction. Compressing in-plane Bi-S distance enhances Bi-S orbital overlaps, and bulk superconductivity emerges. Similar chemical pressure effect can be produced by Se substitution for the S site, as demonstrated in $LaO_{0.5}F_{0.5}BiS_{2-y}Se_y$ and $Eu_{0.5}La_{0.5}FBiS_{2-y}Se_y$ [23,24]. Although Bi-Ch plane expands by the substitution of in-plane $S^{2-}$ by larger $Se^{2-}$ [25], the packing density of Bi-Ch plane is enhanced due to the almost fixed structure (volume) of the LaO blocking layer [21]. Then, bulk superconductivity is induced by in-plane chemical pressure effect in $LaO_{0.5}F_{0.5}BiS_{2-y}Se_y$ and $Eu_{0.5}La_{0.5}FBiS_{2-y}Se_y$, as well.

Although the in-plane chemical pressure can be a good indicator to qualitatively discuss the emergence of bulk superconductivity in the BiCh$_2$-based compounds, deeper investigation on the essential physical parameters, which are directly correlating with the superconductivity, is needed to fully understand the mechanisms of the emergence of superconductivity in the BiCh$_2$-based superconductors. In this study, we have clarified that the suppression of the in-plane local disorder, which is a typical characteristic of BiCh$_2$-based compounds, is the most essential for the emergence of superconductivity. The in-plane disorder in the BiCh$_2$-based system has been detected using extended X-ray absorption fine structure (EXAFS), X-ray diffraction, and neutron diffraction [21, 26-29]. In addition, we showed that the in-plane disorder could be reduced by the chemical pressure effects [27,29]; the details on the suppression of in-plane disorder in LaO$_{0.5}$F$_{0.5}$BiS$_{2-y}$Se$_y$ and LaO$_{1-x}$F$_x$BiSSe will be discussed later with Fig. 2. Therefore, we have focused on the LaO$_{1-x}$F$_x$BiSSe system, in which in-plane disorder was expected to be suppressed by the chemical pressure effect generated by the Se substitution of the BiCh$_2$ layer, and systematically investigated the crystal structure and physical properties of the LaO$_{1-x}$F$_x$BiSSe system.

Polycrystalline samples of LaO$_{1-x}$F$_x$BiSSe with $x = 0$, 0.05, 0.1, 0.2, 0.3, 0.4, and 0.5 were prepared by a solid state reaction method. Bi$_2$S$_3$ and Bi$_2$Se$_3$ were pre-synthesized by reacting Bi (99.999%), S (99.99%), and Se (99.999%) grains. Powders of Bi$_2$O$_3$ (99.99%), La$_2$S$_3$ (99.9%), BiF$_3$ (99.9%), Bi$_2$S$_3$, and Bi$_2$Se$_3$ powders and Bi (99.999%) grains were mixed, pelletized, sealed into an evacuated quartz tube, and heated at 700 ℃ for 20 h. The obtained sample was ground, mixed, pelletized, and heated with the same heating condition.

The crystal structure of the samples and impurity phases were determined by powder synchrotron X-ray diffraction (XRD) with energy of 25 keV ($\lambda = 0.496574$ Å) at the beamline BL02B2 of SPring-8 under a proposal No. 2016B1078 for $x = 0$, 0.1, 0.2, 0.3, 0.4, and 0.5. The synchrotron XRD experiments were performed with a sample rotator system at room temperature, and the diffraction data were collected using a high-resolution one-dimensional semiconductor detector MYTHEN with a step of $2\theta = 0.006°$. The crystal structure parameters were refined using the Rietveld method with RIETAN-FP [30]. For the in-plane Bi and Ch1 site (see Fig. 1(c) for definitions of Ch sites), anisotropic displacement parameters $U_{11}$ and $U_{33}$ were used in the Rietveld analyses. For $x = 0$ and 0.1, $U_{22}$ was fixed as $U_{11}$. Using the refined structure parameter, schematic image of the structure was drawn using VESTA [31]. For, $x = 0.05$, we have checked the phase purity by conventional XRD with a Cu-$K\alpha$ radiation.

Temperature dependence of electrical resistivity was measured using a four-terminal method. For $x = 0.05$ with a $T_c$ less than 2 K, the resistivity measurements were performed with the $^3$He system of Physical Property Measurement System (PPMS, Quantum Design) down to $T = 0.5$ K. Temperature dependence of magnetic susceptibility was measured using a superconducting quantum interference devise (SQUID) magnetometer with Magnetic Property Measurement System (MPMS-3) by the SQUID-VSM mode. Seebeck coefficient at room temperature was measured using a ZEM-3 system (Advance RIKO).

Figure 1(a) shows the synchrotron XRD patterns for $x = 0$–0.5. We observed peak shifts according to the increase in F concentration in LaO$_{1-x}$F$_x$BiSSe. In addition, for $x = 0$, we observed a peak splitting of typical ($h$00) peaks, which indicates that the structure distorted into monoclinic. The lowering symmetry in $x = 0$ was unexpected, because, in a previous study [32], we refined the synchrotron XRD pattern for LaOBiSSe ($x = 0$) using a tetragonal ($P4/nmm$) structural model. In the present experiment, however, we could detect this small peak splitting in $x = 0$ by using the high-resolution detector. Figure 1(b) shows the zoomed profiles at around the (200) peak. For $x = 0$, the (200) peak of the tetragonal structure split into (200) and (020) of the monoclinic structure. For x = 0.1, the corresponding peak is still broad, as compared to those of $x = 0.2$–0.5, indicating that the structure of $x = 0.1$ is monoclinic rather than tetragonal. To refine the crystal structure parameters, Rietveld analyses were performed for all the XRD patterns (See Table S1 and S2 and Fig. S1 in Supplemental Materials [33].). For $x = 0$ and 0.1, the XRD patterns were refined using a monoclinic space group of ($P2_1/m$), which is the same as the high-pressure phase of LaO$_{0.5}$F$_{0.5}$BiS$_2$ [10] and single-crystal LaOBiS$_2$ [34]. The refined lattice constant of $\beta$ is 90.189(2) for $x = 0$ and 90.101(1) deg. for $x = 0.1$. For $x = 0.2$–0.5, the splitting of the (200) peak is suppressed, and the Rietveld analyses with a tetragonal space group of $P4/nmm$ resulted in good refinements.

Figure 1(d) and 1(e) show the F concentration dependence lattice constants of $a$, $b$, and $c$. According to the phase transition from monoclinic to tetragonal, $a$ and $b$ of $x = 0.1$ approaches each other. In the tetragonal structure, $\underline{a}$ increases with increasing $x$. In contrast, $c$ decreases with increasing $x$. These evolutions of lattice constants should relate to the changes in the Bi valence state and the inter-layer bonding states. Figure 1(f) shows the F concentration dependences of the Bi-Ch distances. Although the in-plane Bi-Ch1 distance and the Bi-Ch2 distance does not largely change by F substitution, the out-of-plane Bi-Ch1 distance monotonically decreases with increasing $x$. In addition, as listed in Table S1, the Se occupancy at the in-plane Ch1 site is almost the same for $x = 0.1$–$0.5$, indicating that the local structure of the in-plane sites does not change. In Ref. 21, we proposed that the in-plane chemical pressure was a good indicator of the emergence of bulk superconductivity in BiCh$_2$-based compounds and $T_c$. In addition, the Se substitution for the S site of the BiS$_2$ conducting layer is one of the most effective method to enhance in-plane chemical pressure, as revealed in LaO$_{0.5}$F$_{0.5}$BiS$_{2-y}$Se$_y$ and Eu$_{0.5}$La$_{0.5}$FBiS$_{2-y}$Se$_y$. On the basis of the in-plane chemical pressure scenario, we could expect that the BiSSe layer of the LaO$_{1-x}$F$_x$BiSSe system should be suitable for discussing the emergence of superconductivity by carrier doping.

To further clarify the essence of the in-plane chemical pressure effect in the REOBiCh$_2$ system, we have performed Rietveld refinements for LaO$_{0.5}$F$_{0.5}$BiS$_{2-y}$Se$_y$ [21,23] and LaO$_{1-x}$F$_x$BiSSe using anisotropic displacement parameters of in-plane Bi and Ch1. See Supplemental Materials (Table S1) for all displacement parameters of LaO$_{1-x}$F$_x$BiSSe. Figure 2(a) shows the Se concentration dependences of $U_{11}$ for Bi and Ch1 of LaO$_{0.5}$F$_{0.5}$BiS$_{2-y}$Se$_y$. As reported in Ref. 23, the boundary between bulk superconductor phases (Bulk SC in the figure) and weak (filamentary) superconductor phases (Weak SC in the figure) is around $x = 0.4$ in LaO$_{0.5}$F$_{0.5}$BiS$_{2-y}$Se$_y$. We notice that, for $x \leq 0.4$, $U_{11}$ for Ch1 is apparently larger than that of Bi. If the Bi and Ch1 atoms, which forms the Bi-Ch conducting plane, are vibrating with a close amplitude, the values of $U_{11}$ for Bi and Ch1 should be close. However, $U_{11}$ for Ch1 does not correspond to $U_{11}$ of Bi, indicating that there is a factor other than lattice vibration. The factor, which is the cause of large $U_{11}$ for Ch1, is *local in-plane disorder*. In the REOBiS$_2$-based (and EuFBiS$_2$-based) systems, in-plane disorder has been observed by extended X-ray absorption fine structure (EXAFS), X-ray diffraction, and neutron diffraction [21, 26-29]. The local displacements of Ch1 locally lowers the structural symmetry of the Bi-Ch plane from a square to a zig-zag chain. The distorted local structure of the Bi-Ch1 bond is the same as that of monoclinic phase, observed in the high pressure phase of LaO$_{0.5}$F$_{0.5}$BiS$_2$ [10] or single crystals of LaOBiS$_2$ [34]. Furthermore, we have observed that the in-plane disorder could be reduced by the chemical pressure effects [27]. As shown in Fig. 2(a), $U_{11}$ for Ch1 decreases with increasing Se concentration and becomes almost the same values to $U_{11}$ for Bi in LaO$_{0.5}$F$_{0.5}$BiS$_{2-y}$Se$_y$, which indicates that the in-plane disorder is suppressed by Se substitution. Then, bulk superconductivity is induced in this region with less in-plane disorder. This behavior is quite similar to those observed in a cousin system Eu$_{0.5}$La$_{0.5}$FBiS$_{2-y}$Se$_y$ [29].

Figure 2(b) shows the F concentration dependences of $U_{11}$ for Bi and Ch1 of LaO$_{1-x}$F$_x$BiSSe. In contrast to LaO$_{0.5}$F$_{0.5}$BiS$_{2-y}$Se$_y$, $U_{11}$ for Ch1 is small and almost corresponds to $U_{11}$ for Bi, indicating that the in-plane disorder does not exist and $U_{11}$ parameters reflect the amplitude of lattice vibration only. Therefore, from the viewpoint of local disorder, the LaO$_{1-x}$F$_x$BiSSe system is an ideal system to discuss the intrinsic nature of the emergence of superconductivity in the BiCh$_2$ layers by carrier doping. Indeed, bulk superconductor phases suddenly appear at $x = 0.1$, and all samples with $x \geq 0.1$ shows bulk superconductivity, as will be shown next.

Figure 3 shows the temperature dependences of magnetic susceptibility of LaO$_{1-x}$F$_x$BiSSe after ZFC and FC. Even at $x = 0.1$, a sharp superconducting transition was observed, which indicates the emergence of bulk superconductivity. The irreversible temperature $T_{irr}$, which was defined as the temperature where the ZFC curve deviated from the FC curve, is $T_{irr} = 3.3$ K for $x = 0.1$. For $x = 0.2$–0.5, sharp superconducting transitions were observed, and $T_{irr}$ was 3.8, 3.9, 3.9, and 3.9 K for $x = 0.2$, 0.3, 0.4, and 0.5, respectively. Interestingly, $T_{irr}$ does not largely change for $x = 0.2$–0.5, in spite of the crystal structure evolution by the F substitution (Fig. 1).

Figure 4(a) shows the temperature dependences of electrical resistivity for $x = 0$, 0.05, 0.1, 0.2, 0.3, 0.4, and 0.5. For $x = 0$, the temperature dependence of resistivity shows an upturn. This behavior is suddenly suppressed with 5% F substitution. Metallic behavior is observed for $x \geq 0.05$. For typical REO$_{1-x}$F$_x$BiS$_2$ [2,16-19,35], a metallic behavior is not observed in resistivity-temperature ($\rho$-$T$) characteristics, and a semiconducting-like behavior is observed, in spite of carrier doping by F substitutions. In contrast, for LaO$_{1-x}$F$_x$BiSSe, a metallic behavior is observed in the $\rho$-$T$ measurements for all electron-doped samples ($x = 0.05$–0.5). These results suggest that the metallic conductivity can be induced by electron doping in BiCh$_2$-based compounds with less in-plane disorder, as expected from the band structure [7,8].

For all electron doped samples, a sharp superconducting transition was observed. On the transition temperature, we estimated $T_c^{zero}$ as the temperature where resistivity becomes zero. For $x = 0.05$, a superconducting transition with $T_c^{zero} = 1.5$ K was observed. Although we could not measure the temperature dependence of magnetic susceptibility for $x = 0.05$, the sharp superconducting transition in $\rho$-$T$ implies the emergence of homogeneous superconducting states in the $x = 0.05$ sample. For $x = 0.1$, $T_c^{zero}$ is 3.2 K. For other samples ($x = 0.2$–0.5), $T_c$ is around 3.8 K. According to the band calculations, the Fermi surface morphology changes by F substitution [7,8]. In addition, it has been experimentally confirmed by angle-resolved photoemisison spectroscopy (ARPES) [36-38]. In addition, for same systems, the actual carrier concentration determined from ARPES differs from that expected from the F concentration. Including this peculiar phenomena, the relationship between carrier doping and the superconductivity has not been fully understood in the $BiCh_2$-based compounds. Therefore, the fact that $T_c^{zero} \sim 3.8$ K was observed for $x = 0.2$–0.5 seems quite interesting. From the crystal structure analyses of $LaO_{1-x}F_xBiSSe$, we have observed clear changes of the lattice constants and the inter-plane distance (Bi-Ch1(out-of-plane)), which suggests that the actual F concentration changes between $x = 0.2$ and 0.5 (nominal).

To discuss the change in actual carrier concentration by the F substitution, the Seebeck coefficient ($S$) at room temperature ($\sim$295 K) and $\sim$560 K were measured. In addition, we calculated $S$ from the electronic structure of this system. We calculated the electronic structure in LaOBiSSe by first-principles band calculations (WIEN2k package [39]). We then took 1000 k-points and $RK_{max} = 7$, where R is the muffin-tin radius and $K_{max}$ is the maximum value of the reciprocal lattice vectors, and used the experimentally determined lattice structure in Ref. 40. The Seebeck coefficient was calculated using BoltzTraP package [41] with 100000 k-points.

Figure 5(a) shows the nominal $x$ (F concentration) dependences of Seebeck coefficient measured at $T \sim 295$ and $\sim$560 K. For all samples, negative $S$ was observed. Figure 5(b) shows the electron concentration dependences of calculated $S$ for the LaOBiSSe-based structure. Comparing the experimental data and calculated results of $S$, we found that the large increase in $S$ by 10% electron doping seems the same trend. On the experimental results, $S$ slightly increases from $x = 0.1$ to $x = 0.5$. On the calculated $S$ for $T = 300$ K, a flat region shows up at around $x = 0.2$–0.3, and the difference in $S$ between $x = 0.15$ and 0.35 is only $\Delta S = 10$ μV/K. Then, $S$ begins to increase at $x \geq 0.35$. Namely, the trend of the carrier concentration dependences of $S$ in the experiments and the calculations does not correspond at higher $x$. On the basis of these results, we consider that, at higher $x$, the actual concentration is smaller than the nominal values, while it actually increases with increasing nominal $x$. In contrast, at $x = 0.1$–0.2, the actual value would be larger than the nominal value because even $x = 0.1$ gets on the plateau region, which should begin at around $x = 0.15$, on the basis of the calculations. For the high-temperature data, the $S$ (experimental) clearly changes by F substitution between $x = 0.1$–0.4, which indicates that the actual carrier concentration evidently changes by the F substitution in this region, which is consistent with the monotonic shift of lattice constant.

From the discussion above, we consider that the electron carrier concentration obviously changes between $x$ (nominal) = 0–0.5, but the actual electron concentration may be different from the nominal values ($x$). We speculate that the actual electron concentration for the $T_c$-flat region (nominal $x$ =0.2–0.5) is within 0.15–0.35.

Finally, we have established a superconductivity phase diagram of LaO$_{1-x}$F$_x$BiSSe with information about the crystal structure and the normal-state transport characteristics. In an ideal BiCh$_2$-based compound with less in-plane disorder, metallic transport and superconductivity are suddenly induced by a small amount of electron doping. For LaO$_{1-x}$F$_x$BiSSe, superconductivity is induced in both monoclinic and tetragonal structures. In addition, the electron-doping dependence of $T_c$ exhibits an anomalously flat phase diagram. The intrinsic superconductivity phase diagram of the BiCh$_2$-based compounds should be useful for understanding of the mechanisms of superconductivity in the system.

In conclusion, we have investigated the crystal structure and the physical properties of LaO$_{1-x}$F$_x$BiSSe to reveal the intrinsic superconductivity phase diagram of the BiCh$_2$-based compounds. From synchrotron X-ray diffraction and Rietveld refinements with anisotropic displacement parameters for LaO$_{1-x}$F$_x$BiSSe and LaO$_{0.5}$F$_{0.5}$BiS$_{2-y}$Se$_y$, we revealed that the essence of in-plane chemical pressure effect was the suppression of the local in-plane disorder in the BiCh$_2$ layers. The in-plane disorder causes the absence of metallic conductivity and bulk superconductivity in typical REO$_{0.5}$F$_{0.5}$BiS$_2$. Since the in-plane disorder was fully suppressed for all $x$ in LaO$_{1-x}$F$_x$BiSSe, this system is one of the most ideal system to discuss the intrinsic nature of the emergence of superconductivity in the BiCh$_2$-based compounds. In LaO$_{1-x}$F$_x$BiSSe, metallic conductivity and superconductivity are suddenly induced by a small amount of electron doping. For LaO$_{1-x}$F$_x$BiSSe, superconductivity is induced in both monoclinic and tetragonal structures. In addition, the electron-doping dependence of $T_c$ exhibits an anomalously flat phase diagram.

**Acknowledgment**

The authors thank Y. Goto (Tokyo University), K. Terashima (Okayama University), and T. Katase (Hokkaido University) for fruitful discussion. This work was partly supported by

Grants-in-Aid for Scientific Research (Nos. 15H05886, 15H05884, 25707031, 15H03693, 16F16028 and 16H04493).

*E-mail: mizugu@tmu.ac.jp

1) Y. Mizuguchi, H. Fujihisa, Y. Gotoh, K. Suzuki, H. Usui, K. Kuroki, S. Demura, Y. Takano, H. Izawa, and O. Miura, Phys. Rev. B 86, 220510 (2012).

2) Y. Mizuguchi, S. Demura, K. Deguchi, Y. Takano, H. Fujihisa, Y. Gotoh, H. Izawa, and O. Miura , J. Phys. Soc. Jpn. 81, 114725 (2012).

3) Y. Mizuguchi, J. Phys. Chem. Solids 84, 34 (2015).

4) J. B. Bednorz and K. Müller, Z. Phys. B 64, 189 (1986).

5) Y. Kamihara, T. Watanabe, M. Hirano, and H. Hosono, J. Am. Chem. Soc. 130, 3296 (2008).

6) C. Morice, R. Akashi, T. Koretsune, S. S. Saxena, and R. Arita arXiv:1701.02909.

7) H. Usui, K. Suzuki, and K. Kuroki, Phys. Rev. B 86, 220501 (2012).

8) H. Usui and K. Kuroki, Nov. Supercond. Mater. 1, 50 (2015).

9) R. Jha, H. Kishan, and V. P. S. Awana, J. Phys. Chem. Solids 84, 17 (2015).

10) T. Tomita, M. Ebata, H. Soeda, H. Takahashi, H. Fujihisa, Y. Gotoh, Y. Mizuguchi, H. Izawa, O. Miura, S. Demura, K. Deguchi, and Y. Takano, J. Phys. Soc. Jpn. 83, 063704 (2014).

11) C. T. Wolowiec, D. Yazici, B. D. White, K. Huang and M. B. Maple, Phys. Rev. B 88, 064503 (2013).

12) C. T. Wolowiec, B. D. White, I. Jeon, D. Yazici, K. Huang, and M. B. Maple, J. Phys.: Condens. Matter 25, 422201 (2013).

13) K. Deguchi, Y. Mizuguchi, S. Demura, H. Hara, T. Watanabe, S. J. Denholme, M. Fujioka, H. Okazaki, T. Ozaki, H. Takeya, T. Yamaguchi, O. Miura, and Y. Takano, EPL 101, 17004 (2013).

14) J. Kajitani, K. Deguchi, A. Omachi, T. Hiroi, Y. Takano, H. Takatsu, H. Kadowaki, O. Miura, and Y. Mizuguchi, Solid State Commun. 181, 1 (2014).

15) J. Kajitani, K. Deguchi, T. Hiroi, A. Omachi, S. Demura, Y. Takano, O. Miura, and Y. Mizuguchi, J. Phys. Soc. Jpn. 83, 065002 (2014).

16) S. Demura, Y. Mizuguchi, K. Deguchi, H. Okazaki, H. Hara, T. Watanabe, S. J. Denholme, M. Fujioka, T. Ozaki, H. Fujihisa, Y. Gotoh, O. Miura, T. Yamaguchi, H. Takeya, and Y. Takano, J. Phys. Soc. Jpn. 82, 033708 (2013).

17) D. Yazici, K. Huang, B. D. White, A. H. Chang, A. J. Friedman, and M. B. Maple, Philosophical Magazine 93, 673 (2012).

18) R. Jha, A. Kumar, S. K. Singh, and V. P. S. Awana, J. Appl. Phys. 113, 056102 (2013).

19) S. Demura, K. Deguchi, Y. Mizuguchi, K. Sato, R. Honjyo, A. Yamashita, T. Yamaki, H. Hara, T. Watanabe, S. J. Denholme, M. Fujioka, H. Okazaki, T. Ozaki, O. Miura, T. Yamaguchi, H. Takeya, and Y. Takano, J. Phys. Soc. Jpn. 84, 024709 (2015).

20) S. Demura, Nov. Supercond. Mater. 2, 1 (2016).

21) Y. Mizuguchi, A. Miura, J. Kajitani, T. Hiroi, O. Miura, K. Tadanaga, N. Kumada, E. Magome, C. Moriyoshi, and Y. Kuroiwa, Sci. Rep. 5, 14968 (2015).

22) J. Kajitani, T. Hiroi, A. Omachi, O. Miura, and Y. Mizuguchi, J. Phys. Soc. Jpn. 84, 044712 (2015).

23) T. Hiroi, J. Kajitani, A. Omachi, O. Miura, and Y. Mizuguchi, J. Phys. Soc. Jpn 84, 024723 (2015).

24) G. Jinno, R. Jha, A. Yamada, R. Higashinaka, T. D. Matsuda, Y. Aoki, M. Nagao, O. Miura, and Y. Mizuguchi, J. Phys. Soc. Jpn. 85, 124708 (2016).

25) M. Tanaka, T. Yamaki, Y. Matsushita, M. Fujioka, S. J. Denholme, T. Yamaguchi, H. Takeya, and Y. Takano, Appl. Phys. Lett. 106, 11260 (2015).

26) E. Paris, B. Joseph, A. Iadecola, T. Sugimoto, L. Olivi, S. Demura, Y. Mizuguchi, Y. Takano, T. Mizokawa, and N. L. Saini, J. Phys.: Condens. Matter 26, 435701 (2014).

27) Y. Mizuguchi, E. Paris, T. Sugimoto, A. Iadecola, J. Kajitani, O. Miura, T. Mizokawa, and N. L. Saini, Phys. Chem. Chem. Phys. 17, 22090 (2015).

28) A. Athauda, J. Yang, S. Lee, Y. Mizuguchi, K. Deguchi, Y. Takano, O. Miura, and D. Louca, Phys. Rev. B 91, 144112 (2014).

29) K. Nagasaka, G. Jinno, O. Miura, A. Miura, C. Moriyoshi, Y. Kuroiwa, and Y. Mizuguchi, arXiv:1701.07575.

30) F. Izumi and K. Momma, Solid State Phenom. 130, 15 (2007).

31) K. Momma and F. Izumi, J. Appl. Crystallogr. 41, 653 (2008).

32) Y. Mizuguchi, A. Miura, A. Nishida, O. Miura, K. Tadanaga, N. Kumada, C. H. Lee, E. Magome, C. Moriyoshi, and Y. Kuroiwa, J. Appl. Phys. 119, 155103 (2016).

33) (Supplemental materials) [Crystal structure parameters, XRD patterns and Rietveld fitting] are provided online.

34) R. Sagayama, H. Sagayama, R. Kumai, Y. Murakami, T. Asano, J. Kajitani, R. Higashinaka, T. D. Matsuda, and Y. Aoki, J. Phys. Soc. Jpn. 84, 123703 (2015).

35) Jie Xing, Sheng Li, Xiaxing Ding, Huang Yang, and Hai-Hu Wen, Phys. Rev. B 86, 214518 (2012).

36) K. Terashima, J. Sonoyama, T. Wakita, M. Sunagawa, K. Ono, H. Kumigashira, T. Muro, M. Nagao, S. Watauchi, I. Tanaka, H. Okazaki, Y. Takano, O. Miura, Y. Mizuguchi, H. Usui, K. Suzuki, K. Kuroki, Y. Muraoka, and T. Yokoya, Phys. Rev. B 90, 220512 (2014).

37) Z. R. Ye, H. F. Yang, D. W. Shen, J. Jiang, X. H. Niu, D. L. Feng, Y. P. Du, X. G. Wan, J. Z. Liu, X. Y. Zhu, H. H. Wen, and M. H. Jiang, Phys. Rev. B 90, 045116 (2014).

38) T. Sugimoto, D. Ootsuki, M. Takahashi, C. Morice, E. Artacho, S. S. Saxena, E. F. Schwier, M. Zheng, Y. Kojima, H. Iwasawa, K. Shimada, M. Arita, H. Namatame, M. Taniguchi, N. L. Saini, T. Asano, T. Nakajima, R. Higashinaka, T. D. Matsuda, Y. Aoki, and T. Mizokawa, Phys. Rev. B 92, 041113 (2015).

39) P. Blaha, K. Schwarz, G. Madsen, D. Kvasnicka, and J. Luitz, WIEN2k: an Augmented Plane Wave Plus Local Orbitals Program for Calculating Crystal Properties; Institute of Physical and Theoretical Chemistry: Vienna, Austria, (2001).

40) M. Tanaka, T. Yamaki, Y. Matsushita, M. Fujioka, S. J. Denholme, T. Yamaguchi, H. Takeya, and Y. Takano, Appl. Phys. Lett. 106, 112601 (2015).

41) G. K.H. Madsen, and D. J. Singh, Comput. Phys. Commun. 175, 67 (2006).

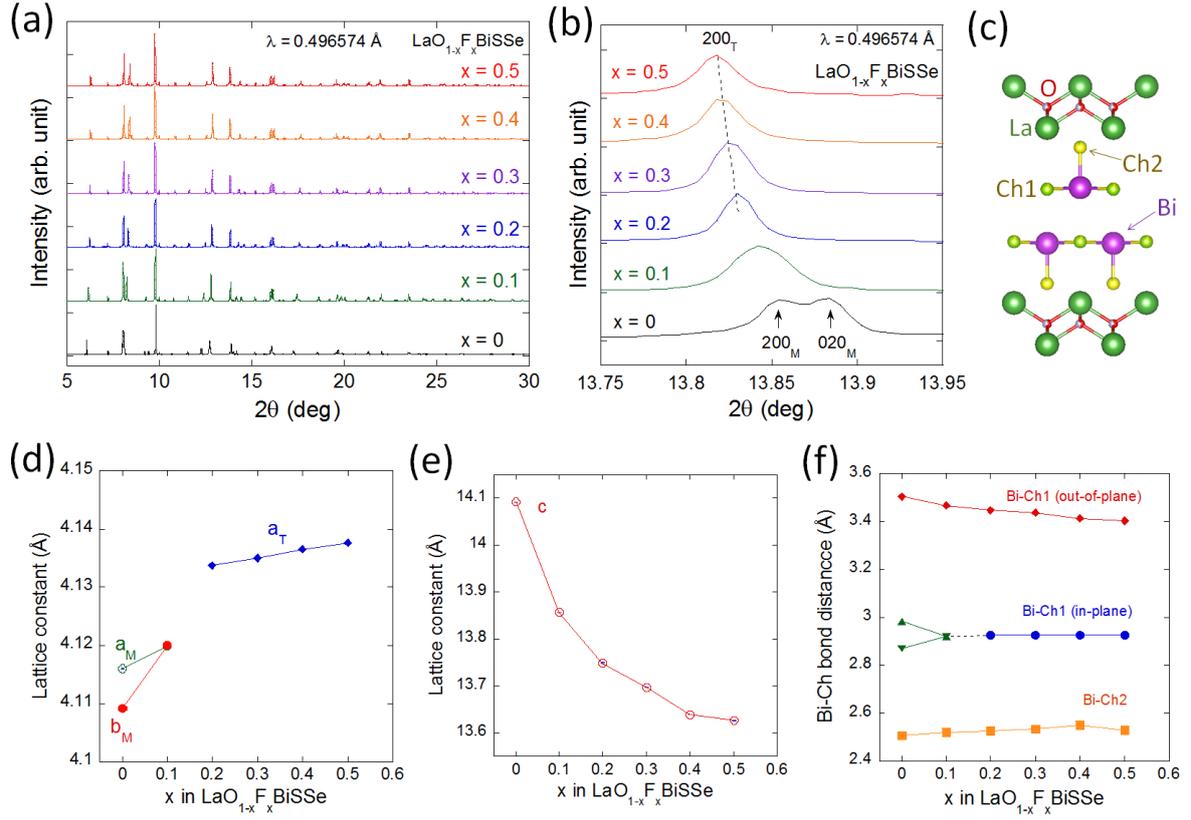

Fig. 1. (a) XRD patterns of LaO$_{1-x}$F$_x$BiSSe with $x$ = 0-1. (b) XRD patterns at around (200) peaks. 200$_T$ denotes the 200 peak of the tetragonal structure, and 200$_M$ and 020$_M$ denote the 200 and 020 peaks of the monoclinic structure. (c) Schematic image of the crystal structure of LaOBiSSe. The chalcogen site in the conducting plane is defined as Ch1, and the other chalcogen site is defined as Ch2. (d) F concentration dependence of lattice constants $a$ and $b$. $a_M$, $b_M$, and $a_T$ denote $a$(monoclinic), $b$(monoclinic), and $a$(tetragonal). (e) F concentration dependence of lattice constants $c$. (f) F concentration dependences of refined Bi-Ch distances.

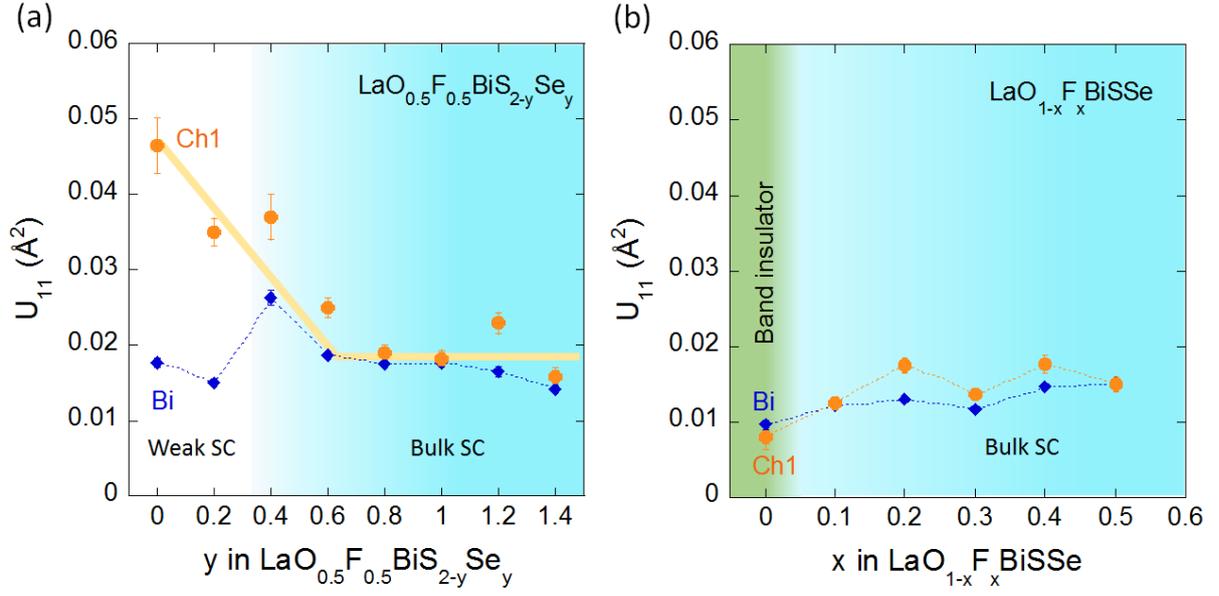

Fig. 2. (a) Se concentration dependences of the anisotropic displacement parameter $U_{11}$ (in-plane component) for the Bi and Ch1 sites for $LaO_{0.5}F_{0.5}BiS_{2-y}Se_y$. (b) F concentration dependences of the anisotropic displacement parameter $U_{11}$ (in-plane component) for the Bi and Ch1 sites for $LaO_{1-x}F_xBiSSe$. Bulk SC denotes bulk superconductor phases, and Weak SC denotes phases in which weak (filamentary) superconducting states are emerging.

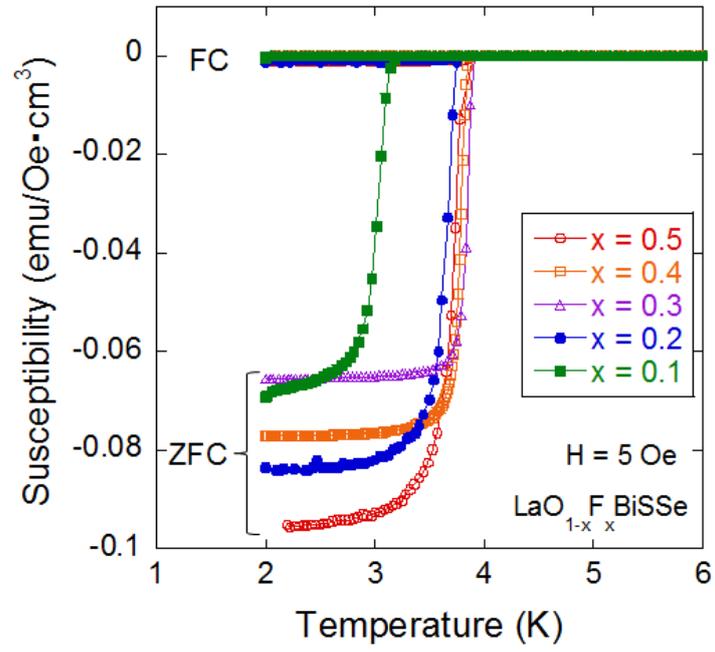

Fig. 3. Temperature dependences of ZFC and FC magnetic susceptibility for $x$ = 0.1–0.5.

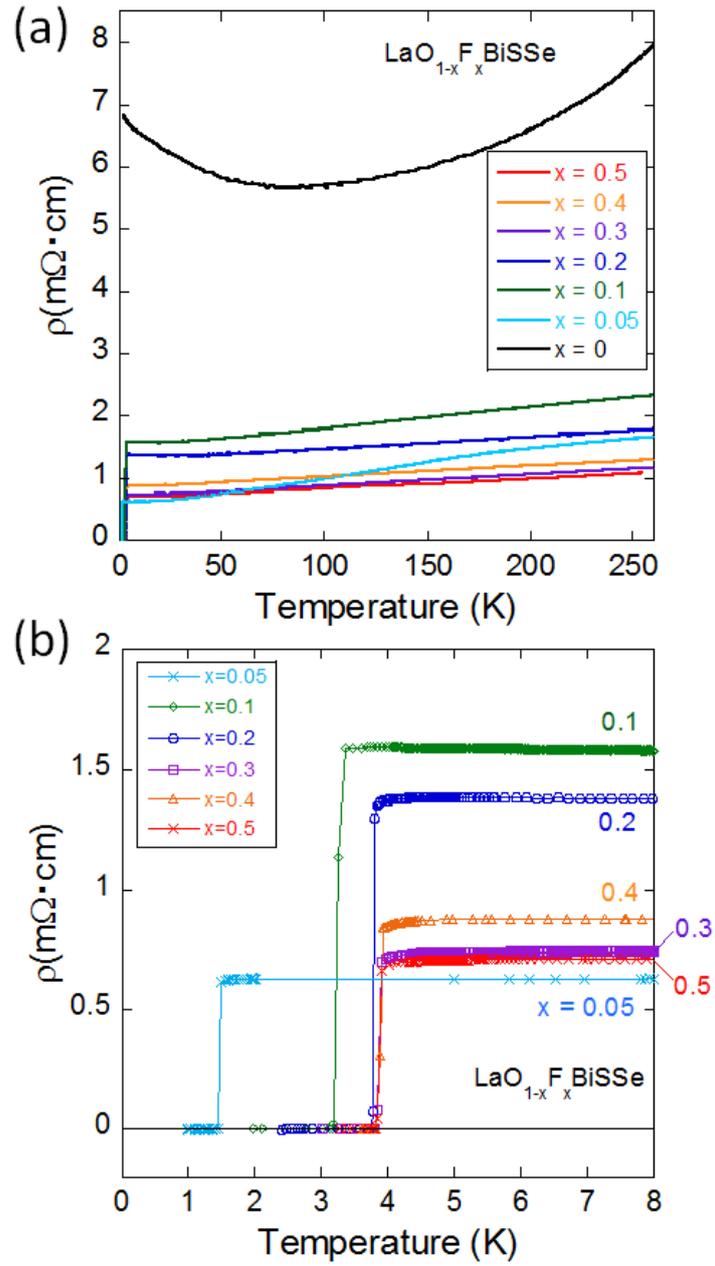

Fig. 4. (a) Fig. 2. Temperature dependences of electrical resistivity for $x$ = 0–0.5. (b) Zoomed figure of temperature dependences of electrical resistivity at around the superconducting transitions for $x$ = 0.1–0.5.

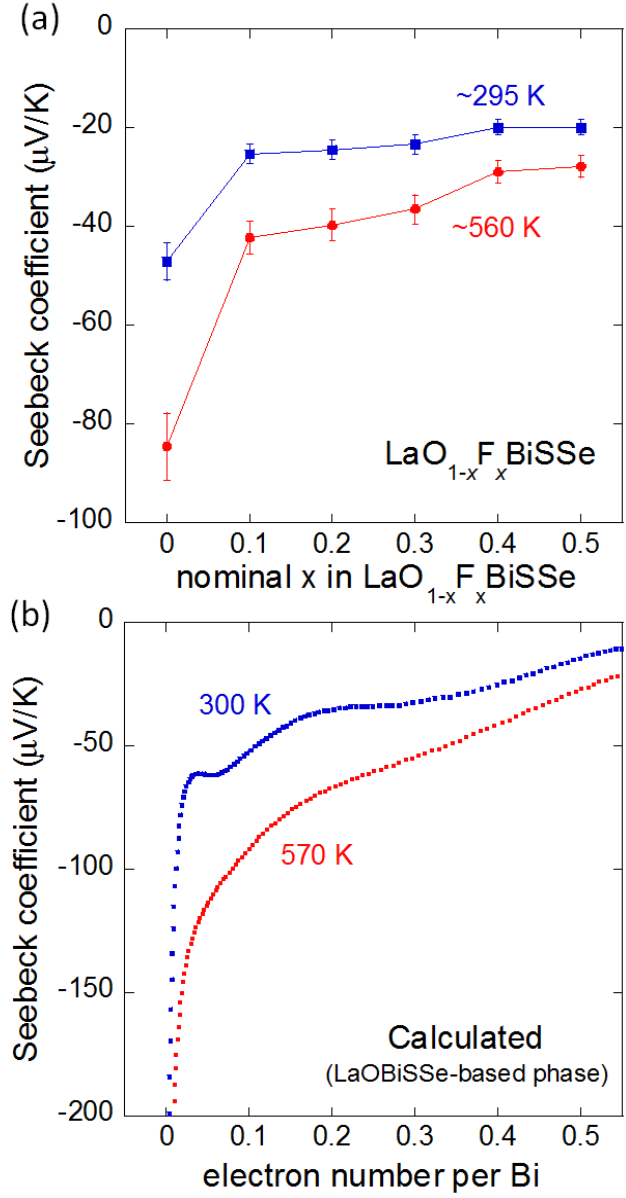

Fig. 5. (a) F concentration dependences of Seebeck coefficient at room temperature (~295 K) and ~560 K for LaO$_{1-x}$F$_x$BiSSe. (b) Calculated Seebeck coefficient at 300 and 570 K for the LaOBiSSe-based phase as a function of electron number per Bi, which corresponds to $x$ in LaO$_{1-x}$F$_x$BiSSe.

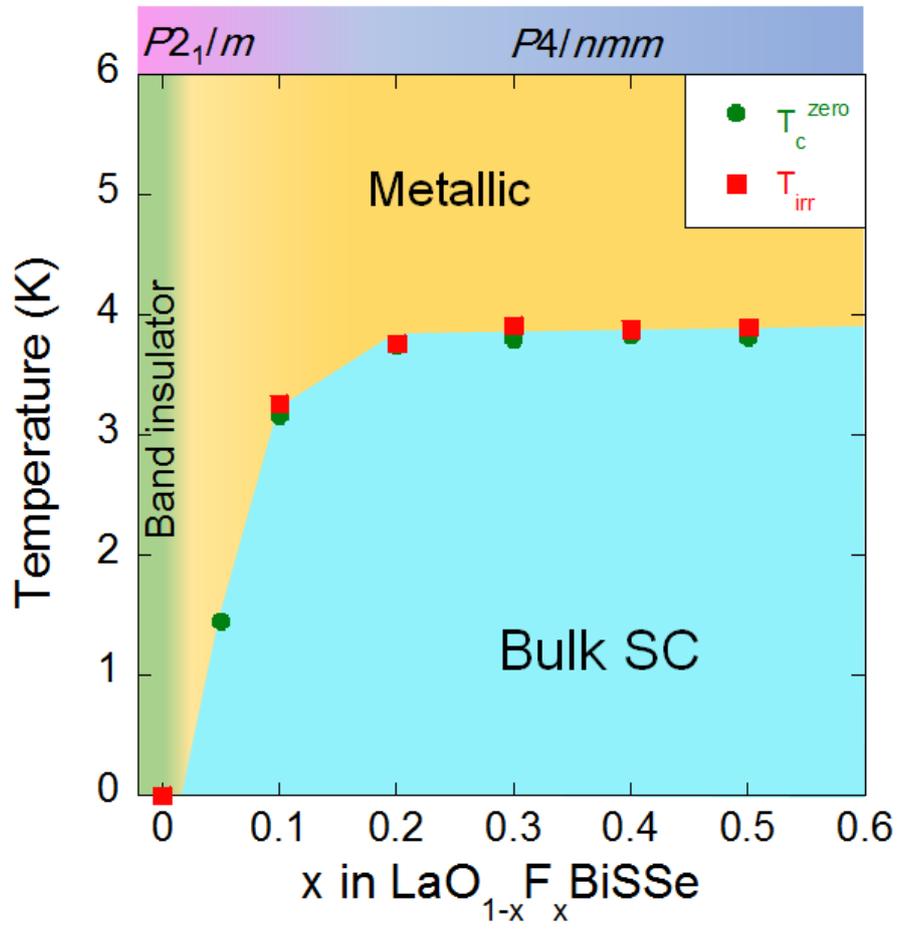

Fig. 6. Superconductivity phase diagram of LaO$_{1-x}$F$_x$BiSSe. Bulk SC denotes bulk superconductor phases. $P2_1/m$ and $P4/nmm$ are space groups of the sample.

Supplemental Materials

# Intrinsic Phase Diagram of Superconductivity in the BiCh₂-based System without In-plane Disorder

Kouhei Nagasaka[1], Atsuhiro Nishida[1], Rajveer Jha[2], Joe Kajitani[2], Osuke Miura[1], Ryuji Higashinaka[2], Tatsuma D. Matsuda[2], Yuji Aoki[2], Akira Miura[3], Chikako Moriyoshi[4], Yoshihiro Kuroiwa[4], Hidetomo Usui[5], Kazuhiko Kuroki[5], Yoshikazu Mizuguchi[1*]

*[1]Department of Electrical and Electronic Engineering, Tokyo Metropolitan University, Hachioji 192-0397, Japan*

*[2]Department of Physics, Tokyo Metropolitan University, Hachioji 192-0397, Japan*

*[3]Faculty of Engineering, Hokkaido University, Kita-13, Nishi-8, Kita-ku, Sapporo, Hokkaido 060-8628, Japan*

*[4]Department of Physical Science, Hiroshima University, 1-3-1 Kagamiyama, Higashihiroshima, Hiroshima 739-8526, Japan*

*[5]Department of Physics, Osaka University, Machikaneyama-cho, Toyonaka 560-0043, Japan*

Table S1. Crystal structure parameters of $LaO_{1-x}F_xBiSSe$ refined using Rietveld fitting.

| $x$ | 0 | 0.1 | 0.2 | 0.3 | 0.4 | 0.5 |
|---|---|---|---|---|---|---|
| space group | $P2_1/m$ | $P2_1/m$ | $P4/nmm$ | $P4/nmm$ | $P4/nmm$ | $P4/nmm$ |
| $a$ (Å) | 4.1160(1) | 4.11986(6) | 4.13377(3) | 4.13497(2) | 4.13661(4) | 4.13767(4) |
| $b$ (Å) | 4.10923(8) | 4.12003(4) | - | - | - | - |
| $c$ (Å) | 14.0917(3) | 13.85661(7) | 13.7492(1) | 13.69716(8) | 13.6386(2) | 13.6260(2) |
| $\beta$ (deg) | 90.189(2) | 90.101(1) | - | - | - | - |
| $U_{11}$(Bi) (Å²) | 0.0096(6) | 0.0121(2) | 0.0130(4) | 0.0116(3) | 0.0146(5) | 0.0152(4) |
| $U_{33}$(Bi) (Å²) | 0.024(1) | 0.0119(2) | 0.0149(6) | 0.0199(5) | 0.0306(8) | 0.0282(8) |
| $U_{11}$(Ch1) (Å²) | 0.0079(17) | 0.0126(4) | 0.0175(10) | 0.0137(7) | 0.018(1) | 0.0150(10) |
| $U_{33}$(Ch1) (Å²) | 0.037(4) | 0.031(1) | 0.032(2) | 0.029(2) | 0.031(3) | 0.029(2) |
| $R_{wp}$ (%) | 10.2 | 4.4 | 6.8 | 8.1 | 7.5 | 8.1 |
| Se occupancy at Ch1 (%) | 86.4(9) | 92.1(3) | 93.8(6) | 91.1(5) | 92.7(7) | 93.3(7) |

Table S2. Atomic coordinates and isotropic displacement parameters of LaO$_{1-x}$F$_x$BiSSe refined using Rietveld fitting.

| $x = 0$ | $x$ | $y$ | $z$ | $U_{iso}$ (Å$^2$) | $x = 0.1$ | $x$ | $y$ | $z$ | $U_{iso}$ (Å$^2$) |
|---|---|---|---|---|---|---|---|---|---|
| La | 0.245(1) | 0.25 | 0.9125(1) | 0.0031(7) | La | 0.2490(6) | 0.25 | 0.90938(5) | 0.0071(2) |
| Bi | 0.259(1) | 0.25 | 0.3641(1) | - | Bi | 0.2545(5) | 0.25 | 0.36760(4) | - |
| Ch1 | 0.260(3) | 0.25 | 0.6128(3) | - | Ch1 | 0.246(1) | 0.25 | 0.6179(1) | - |
| Ch2 | 0.239(5) | 0.25 | 0.1864(5) | 0.012(2) | Ch2 | 0.249(2) | 0.25 | 0.1857(2) | 0.0107(8) |
| O/F | 0.75(1) | 0.25 | 0.020(2) | 0.013 (fixed) | O/F | 0.744(7) | 0.25 | 0.0116(9) | 0.013 (fixed) |

| $x = 0.2$ | $x$ | $y$ | $z$ | $U_{iso}$ (Å$^2$) | $x = 0.3$ | $x$ | $y$ | $z$ | $U_{iso}$ (Å$^2$) |
|---|---|---|---|---|---|---|---|---|---|
| La | 0 | 0.5 | 0.0930(1) | 0.0081(5) | La | 0 | 0.5 | 0.09421(8) | 0.0087(3) |
| Bi | 0 | 0.5 | 0.63026(9) | - | Bi | 0 | 0.5 | 0.62963(8) | - |
| Ch1 | 0 | 0.5 | 0.3795(2) | - | Ch1 | 0 | 0.5 | 0.3786(2) | - |
| Ch2 | 0 | 0.5 | 0.8141(4) | 0.0074(16) | Ch2 | 0 | 0.5 | 0.8147(3) | 0.011(1) |
| O/F | 0 | 0 | 0 | 0.013 (fixed) | O/F | 0 | 0 | 0 | 0.013 (fixed) |

| $x = 0.4$ | $x$ | $y$ | $z$ | $U_{iso}$ (Å$^2$) | $x = 0.5$ | $x$ | $y$ | $z$ | $U_{iso}$ (Å$^2$) |
|---|---|---|---|---|---|---|---|---|---|
| La | 0 | 0.5 | 0.0956(1) | 0.0079(5) | La | 0 | 0.5 | 0.0960(1) | 0.0077(5) |
| Bi | 0 | 0.5 | 0.6288(1) | - | Bi | 0 | 0.5 | 0.6288(1) | - |
| Ch1 | 0 | 0.5 | 0.3785(2) | - | Ch1 | 0 | 0.5 | 0.3790(2) | - |
| Ch2 | 0 | 0.5 | 0.8157(4) | 0.010(2) | Ch2 | 0 | 0.5 | 0.8145(4) | 0.0070(17) |
| O/F | 0 | 0 | 0 | 0.013 (fixed) | O/F | 0 | 0 | 0 | 0.013 (fixed) |

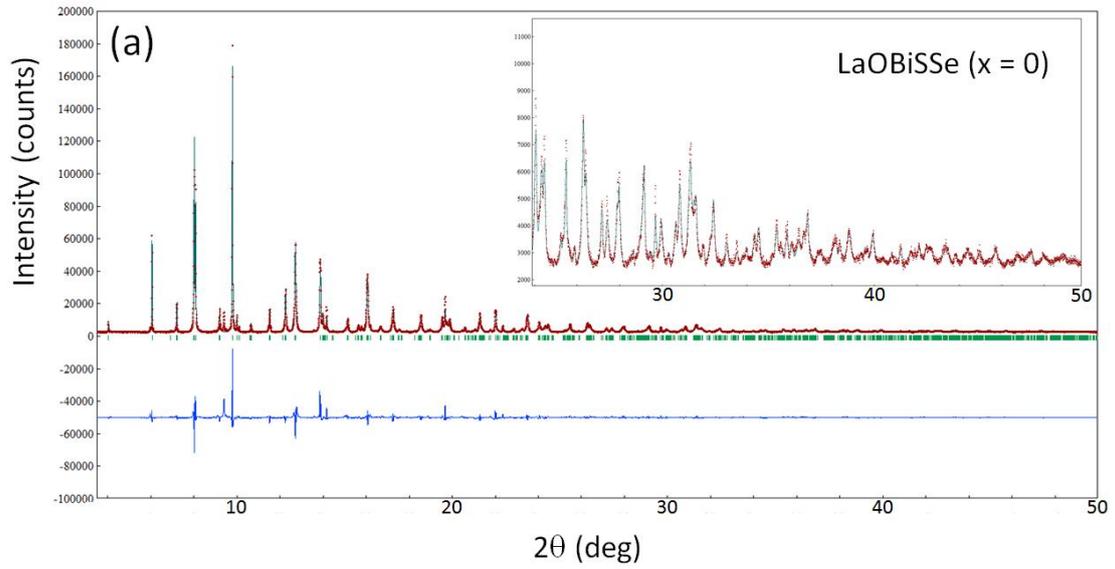

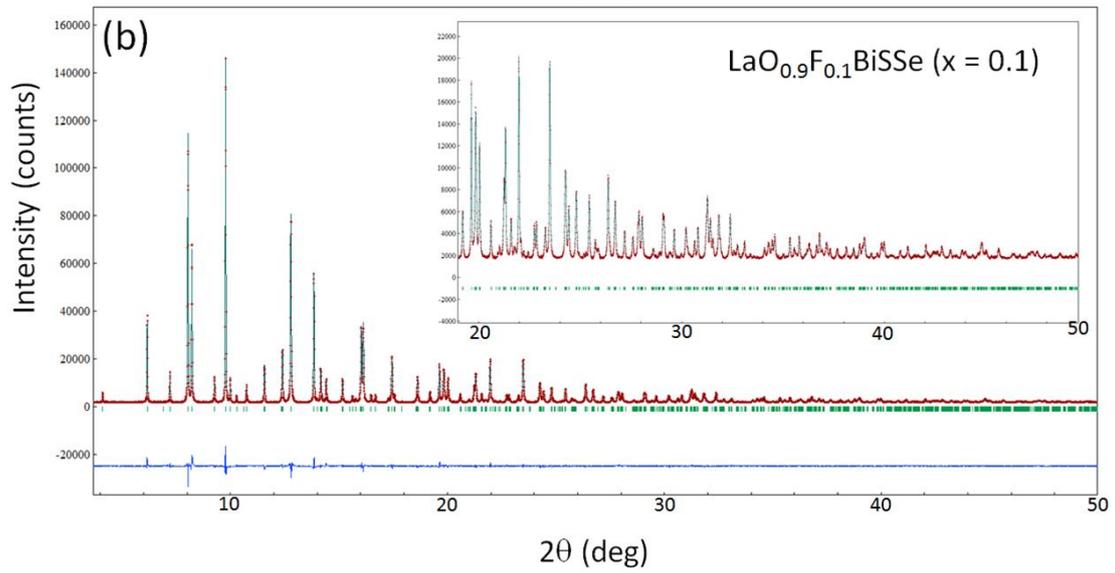

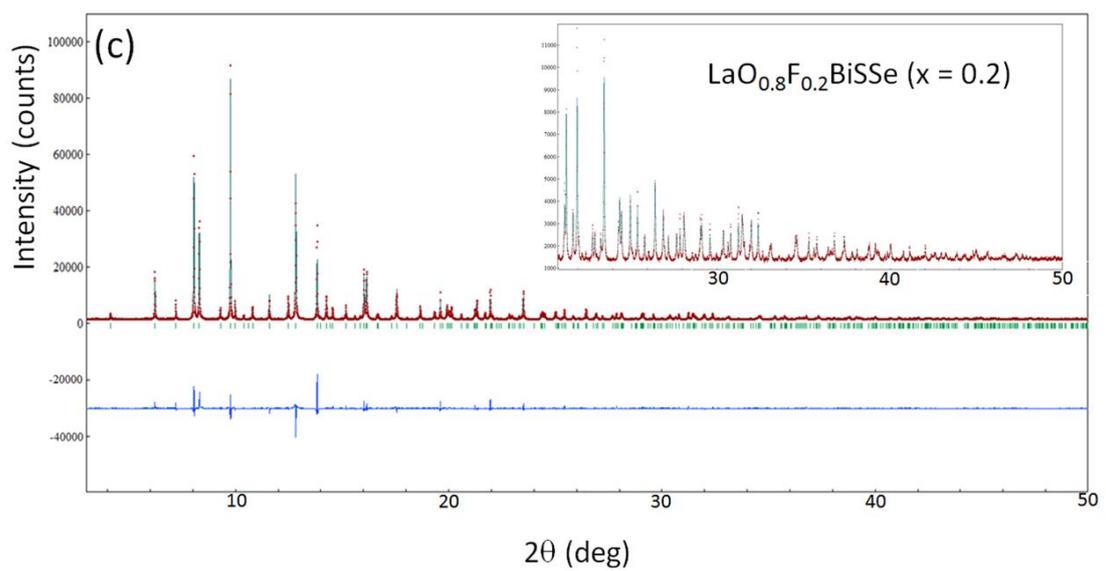

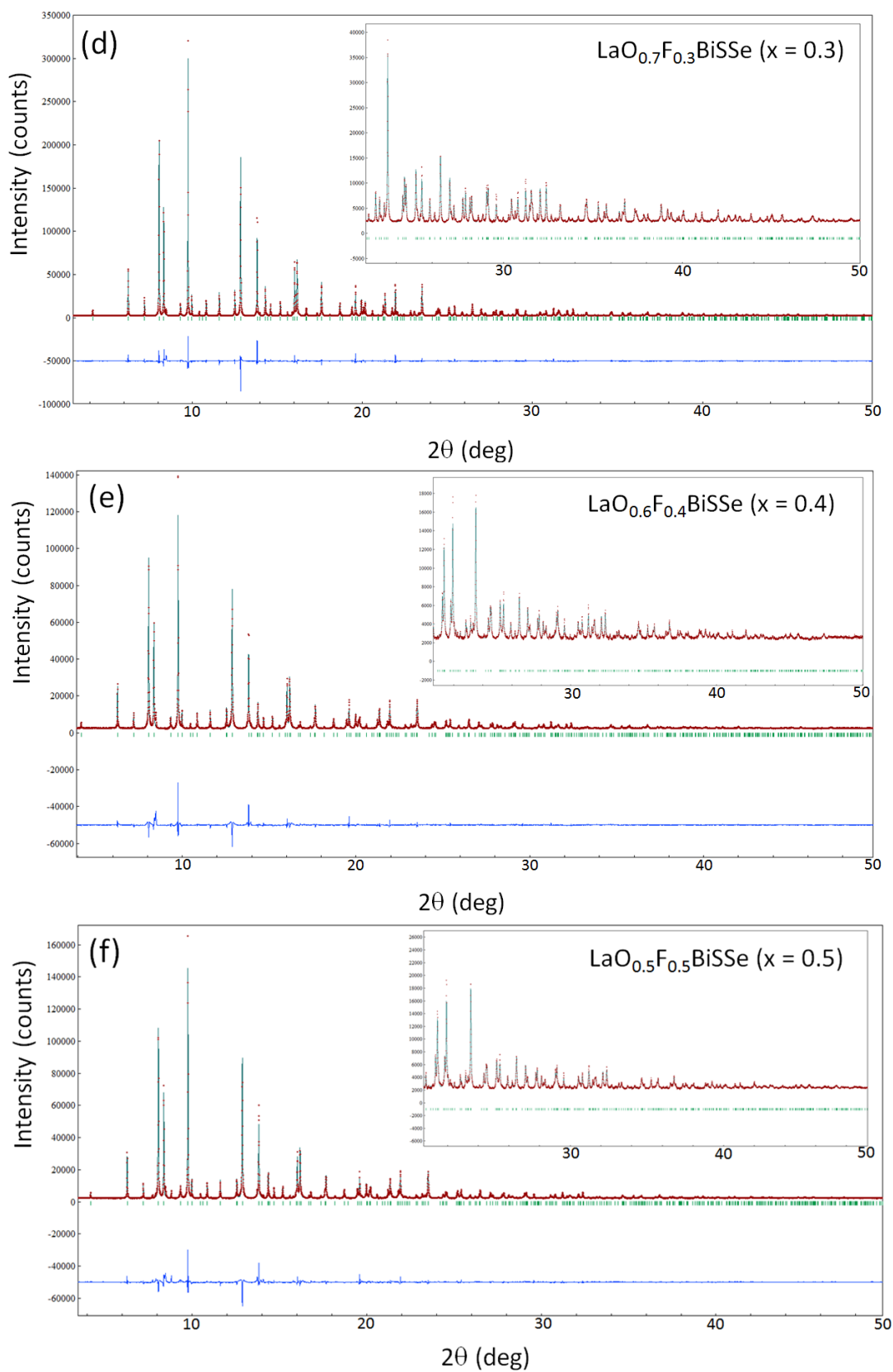

Fig. S1. Synchrotron powder XRD patterns and Rietveld fitting of LaO$_{1-x}$F$_x$BiSSe. Insets shows patterns enlarged at higher angles. The good fitting at higher angles suggests that the assumed structural model is correct. For $x = 0$, small amount of La$_2$O$_3$ impurity was detected.